\documentclass[journal=nalefd,manuscript=letter]{achemso}

\usepackage[version=3]{mhchem} 
\usepackage[T1]{fontenc}       
\usepackage{color}
\setkeys{acs}{articletitle=true}
\setkeys{acs}{maxauthors = 0}

\usepackage{soul}

\email{kanelson@mit.edu}

\author{Jiaojian Shi$^\bigtriangleup$}
\affiliation[MITChem]{Department of Chemistry, Massachusetts Institute of Technology, Cambridge, MA 02139, USA}

\author{Frank Y. Gao$^\bigtriangleup$}
\affiliation[MITChem]{Department of Chemistry, Massachusetts Institute of Technology, Cambridge, MA 02139, USA}

\author{Zhuquan Zhang$^\bigtriangleup$}
\affiliation[MITChem]{Department of Chemistry, Massachusetts Institute of Technology, Cambridge, MA 02139, USA}

\author{Hendrik Utzat}
\affiliation[MITChem]{Department of Chemistry, Massachusetts Institute of Technology, Cambridge, MA 02139, USA}

\author{Ulugbek Barotov}
\affiliation[MITChem]{Department of Chemistry, Massachusetts Institute of Technology, Cambridge, MA 02139, USA}

\author{Ardavan Farahvash}
\affiliation[MITChem]{Department of Chemistry, Massachusetts Institute of Technology, Cambridge, MA 02139, USA}

\author{Jinchi Han}
\affiliation[MITEE]{Department of Electrical Engineering and Computer Science, Massachusetts Institute of Technology, Cambridge, MA 02139, USA}

\author{Jude Deschamps}
\affiliation[MITChem]{Department of Chemistry, Massachusetts Institute of Technology, Cambridge, MA 02139, USA}

\author{Chan-Wook Baik}
\affiliation[Samsung]{Photonic Device Lab, Samsung Advanced Institute of Technology, 16678 Suwon, Republic of Korea}

\author{Kyung Sang Cho}
\affiliation[Samsung]{Photonic Device Lab, Samsung Advanced Institute of Technology, 16678 Suwon, Republic of Korea}

\author{Vladimir Bulovi\'c}
\affiliation[MITEE]{Department of Electrical Engineering and Computer Science, Massachusetts Institute of Technology, Cambridge, MA 02139, USA}

\author{Adam P. Willard}
\affiliation[MITChem]{Department of Chemistry, Massachusetts Institute of Technology, Cambridge, MA 02139, USA}

\author{Edoardo Baldini}
\affiliation[MITPhys]{Department of Physics, Massachusetts Institute of Technology, Cambridge, MA 02139, USA}

\author{Nuh Gedik}
\affiliation[MITPhys]{Department of Physics, Massachusetts Institute of Technology, Cambridge, MA 02139, USA}

\author{Moungi G. Bawendi}
\affiliation[MITEE]{Department of Electrical Engineering and Computer Science, Massachusetts Institute of Technology, Cambridge, MA 02139, USA}

\author{Keith A. Nelson}
\affiliation[MITChem]{Department of Chemistry, Massachusetts Institute of Technology, Cambridge, MA 02139, USA}
\email{kanelson@mit.edu}

\title[]
{Terahertz Field-Induced Reemergence of Quenched Photoluminescence in Quantum Dots}

\abbreviations{}

\begin{document}

%
%
%
%
%

\newpage
\begin{abstract}
Continuous and concerted development of colloidal quantum-dot light-emitting diodes over the past two decades has established them as a bedrock technology for the next generation of displays. However, a fundamental issue that limits the performance of these devices is the quenching of photoluminescence due to excess charges from conductive charge transport layers. Although device designs have leveraged various workarounds, doing so often comes at the cost of limiting efficient charge injection. Here we demonstrate that high-field terahertz (THz) pulses can dramatically brighten quenched QDs on metallic surfaces, an effect which persists for minutes after THz irradiation. This phenomenon is attributed to the ability of the THz field to remove excess charges, thereby reducing trion and non-radiative Auger recombination. Our findings show that THz technologies can be used to suppress and control such undesired non-radiative decay, potentially in a variety of luminescent materials for future device applications.
\end{abstract}
\newpage

In the last decades, the class of zero-dimensional semiconductor crystals dubbed quantum dots (QDs) has revolutionalized the field of optoelectronics.\cite{Murray2000,Klimov2000,Alivisatos1996} Owing to the efficient and tunable light emission of their exciton states, these confined nanosystems represent a superior platform for the development of light emitting diodes (LEDs) to be used in next-generation display technologies.\cite{Shirasaki2013} Achieving a near-unity external quantum efficiency -- defined as the ratio of the number of radiated photons to that of injected carriers -- is the holy grail of QD-based LEDs. However, even in the most advanced device architectures, the external quantum efficiency remains below the limit defined by light extraction efficiency\cite{Mashford2013,Dai2014,Yang2015,Shen2019} and exhibits efficiency droop with increasing current density\cite{Lim2018}. The origin of this efficiency gap lies in part in the quenching of the QD PL caused in a real device by direct contact between the nanocrystals and the conductive electron/hole injection substrates known as charge transport layers (CTLs).\cite{Kuhn1970,Shim2001,Caruge2008,Wood2009,Shirasaki2013} Carriers in the excited states of the QDs can break into separate charges and transfer to CTLs, leaving excess charges behind in the QDs. This leads to the formation of three-body bound states (trions) and promotes non-radiative Auger recombination, thus hindering exciton decay \textit{via} photon emission (Fig.~1a-b)\cite{Kuhn1970,Ford1984,Bae2013}.

To break this efficiency bottleneck, various device designs have been explored with the aim of protecting the QDs from the excess charges. Notable examples include the use of less conductive CTLs\cite{Mashford2013,Yang2015,Shen2019} or the addition of spacing layers\cite{Wood2009,Dai2014}. However, these solutions come at the cost of reduced current injection efficiency\cite{Anikeeva2008}. Another strategy relies on keeping the QDs on top of the CTLs while restoring the PL by means of an external perturbation. One such control parameter is a DC electric field, since it has the potential to remove the excess charges inside the QDs. It has been shown that DC fields at kV/cm level could manipulate charge\cite{Warburton2000}, dissociate and restore excitons\cite{Lundstrom1999} in QDs at cryogenic temperature. Nevertheless, this stimulus failed to brighten the quenched PL on metallic surfaces in previous attempts.\cite{Shimizu2000,Cherniavskaya2003} This is because a weak DC field is insufficient to overcome the potential barrier between the QDs and the conductive surfaces, and thus is unable to discharge the nanocrystals. On the other hand, a high DC field would likely cause dielectric breakdown before the QDs were sufficiently discharged. Therefore, to date, restoring the quenched PL in QD-based LEDs remains a major challenge in optoelectronics.

In this letter, we demonstrate a scheme that accomplishes this goal. This has been made possible by the use of short electromagnetic pulses in the terahertz (THz) spectral range. Free-space THz fields at a sub MV/cm level have been demonstrated to release trapped electrons and contribute to excitonic populations in quantum wells at low temperatures.\cite{Shinokita2013} We adopted field-enhancement structures to increase free-space THz pulses to a several MV/cm level. These single-cycle THz pulses can liberate excess carriers from the QDs without causing dielectric breakdown (Fig.~1c).\cite{Hoffmann2009,Liu2012,Fan2013} We distributed CdSe/CdS core/shell QDs sparsely on a gold THz field enhancement structure (100-$\mu$m wide parallel gold strips separated by 2-$\mu$m wide capacitive gaps\cite{Gray2018}, as shown in Fig.~2a), and monitored the PL intensity by imaging the QD emission under continuous wave (CW) irradiation with 405 nm light. Contact with the gold mimics the case of QDs attached to CTLs, as indicated by the almost complete quench of the PL intensity (Fig.~S3, $t$ $<$ 0).\cite{Zhang2014,Bharadwaj2011} Changes in the PL intensity were monitored (Fig.~1d) following the opening of a shutter which initiated irradiation at a 1-kHz repetition rate by single-cycle THz pulses whose free-space peak field amplitude ($>$170 kV/cm, Fig.~1e-f)\cite{yeh2007generation} was enhanced 20$\times$ by the periodic gold structure. The THz pulses were passed through the silica substrate and the gold field enhancement structure, so the enhanced near-field THz radiation reached the QDs close to the insulating gaps in the structure. Images of QD emission were recorded for many seconds during and after periods of THz irradiation. PL spectra were also recorded with and without THz irradiation under way. PL lifetimes were measured using pulsed 405-nm photoexcitation rather than CW excitation as in the other measurements. All experiments were performed under ambient conditions.

\section{Results and Discussion}

Figure~2b-f show data obtained upon irradiation with 500 kV/cm free-space THz pulses. We observe that a substantial PL signal emerges from multiple regions of the sample and its intensity continues to grow over many seconds. The PL signal is unquenched primarily in those areas with the highest field strength, \textit{e.g.}, the edges of the gold-covered region near the insulating gaps. This suggests that a discharging process, \textit{i.e.}, transfer of excess charges from the QDs back to the gold substrate, is driven by the high THz electric field. We can observe unquenched PL signal even long after the THz irradiation is turned off. Figure~2g displays the time-dependent buildup and decay of THz-assisted PL intensity. This signal was selected from a representative region near one edge of a gold strip (outlined with a dashed red line in Fig.~2f) for a series of four THz irradiation periods, each period lasting 10 s with successive periods separated by 5 minutes. We observe that during the first run the PL response is initially quite weak and gradually increases due to the cumulative effect of thousands of THz pulses. In each subsequent irradiation period, the initial PL intensity is higher, the PL response at the beginning of THz irradiation is more rapid, and the peak PL intensity that is ultimately reached is higher. The peak PL intensity obtained during Run 4 is more than 100 times higher than the quenched PL intensity before THz irradiation. All these features indicate that the QDs do not return to their initial quenched state during the 5-minute intervals between runs. The PL decay after the fourth THz irradiation run, shown in Fig.~2h, is slow and highly non-exponential. After an initial rapid drop over the first 10 s, the decay of the PL slows down and displays a power-law scaling relation with an exponent of 0.22. Similar power-law decays have been observed in disordered systems that undergo charging and recovery processes with non-Markovian transport kinetics.\cite{Scher1975,Monroe1987,Morgan2002} The kinetics indicate that liberated carriers tunnel across a distribution of barriers to recover the initial quenched state. Such a slow recovery process is also consistent with reports on relaxation following the charging of neutral QDs.\cite{Shim2001,Wang2001,Makihara2006}

Mechanistically, the observed PL response can be attributed to the influence of the THz field driving excess free charge carriers on QDs back to the metal. Here, we assume these charge carriers are holes due to the relative work functions of CdSe core, CdS shell, and gold (detailed analysis is provided in Supplementary Note 6).\cite{Mattoussi1998,Zhou2012,Bae2013b,Pal2012} Modeling this influence with a simple three-state kinetic model for the free hole yields qualitative agreement with experimental results (Fig.~2g). In the model, hole transfer between the QD nanocrystal ($Q$) state and metal ($M$) is mediated by a tertiary surface defect ($D$) state. In the absence of THz fields, the energy of free holes on the QD in both the $Q$ and $D$ states is assumed to be below the Fermi level of the metal, giving rise to QD charging and quenched emission rates. When the field is turned on, we assume that excitation with THz pulses increases the energy of QD holes \textit{via} direct charge acceleration, giving rise to hole backtransfer to the metal and PL unquenching. The hysteresis effect that results from successive THz exposures is accounted for \textit{via} an increase in the initial population of the $D$ states in successive runs. Hole transfer between the $Q$ and $D$ states is assumed to be slow compared to transfer between the $D$ and $M$ states, and therefore in the first two runs the PL unquenching is constrained by the low initial population of the $D$ state. Despite the simple description, we find that the observed PL accumulation behaviors are well represented by this kinetic model (see Methods and Supplementary Information for further details).

Figure~3 shows the dependences of THz-induced changes on the PL intensity near the gold edge under differing experimental conditions. As shown in Fig.~3a, the PL change induced by THz excitation is proportional to the optical power density, and no emission is observed without optical light. This confirms that the THz field alone does not drive any emission \textit{via} electrolumincence.\cite{Pein2017} The positive proportionality between the THz induced PL change and the optical excitation flux is consistent with more significant charging due to stronger optical excitation and subsequent charge transfer (more discussion about the origin of charging is included in Fig.~S13 and Supplementary Note~6). We also varied the field strength of the THz pulses illuminating the QD sample, obtaining the results shown in Fig.~3b. We observed a threshold-like behavior for the PL unquenching, with emission only induced by free-space field strengths exceeding 170 kV/cm. Above this field strength, the PL intensity increases roughly linearly with increasing field strength. The high field strengths required to observe the effect (including $\sim$20$\times$ enhancement) explains why previous attempts at liberating carriers from charged QDs with weak DC fields have been unsuccessful. Our results support a mechanism in which the THz field must provide sufficient kinetic energy to the excess carriers to allow them to overcome an energy barrier at the interface between the QDs and the gold layer. Lastly, we investigated the effect of increasing the number of THz pulses incident on the sample, with the overall irradiation period fixed at 2 seconds. The resulting change in THz-assisted PL intensity, plotted in Fig.~3c, was found to be proportional to the THz repetition rate. This result is consistent with the gradual buildup and decay of PL intensity in the different runs of Fig.~2g. While high-field THz excitation may impart sufficient kinetic energy for carriers to leave the QDs, the energy may be dissipated without carrier escape. Multiple excitation events lead to higher cumulative populations of neutral QDs and therefore higher THz-induced PL. There is very little recharging of QDs during the 2-second excitation period.

We also measured the PL intensity from sparse QDs on a metamaterial structure coated with an insulating 50-nm Al$_2$O$_3$ layer. In this case, optically generated excitons in the QDs are protected from charge transfers to gold by the oxide layer. So the QDs stay neutral, and PL emission can be observed from many locations without any THz irradiation. As shown in Fig.~S6, no increased PL intensity is observed upon THz irradiation. In contrast, a slight reduction in the PL intensity occurs, possibly associated with THz-field-induced ionization reported earlier.\cite{Hoffmann2009,Liu2012,Fan2013,Pein2017} We note that the conditions used in this control experiment are identical to those of the runs shown in Fig.~2g, notwithstanding the inherent sample inhomogeneity due to the spin-coating process. By comparing with the PL intensity in this control experiment, we estimate that the maximum PL intensity observed in Run 4 of the THz-assisted unquenching experiment is roughly 50\% of the excitonic PL emission from QDs on insulating layers.

To gain further insight into the effects of THz excitation, we measured the equilibrium PL lifetime on gold and glass using pulsed 405-nm photoexcitation at a 5-MHz repetition rate without THz excitation. We observed substantial acceleration of equilibrium PL decay and redshift of PL spectrum by placing QD on gold, as shown in Fig.~4b and Fig.~S11, consistent with QD charging on metallic surfaces. Also, we measured the PL lifetime in the presence and absence of THz irradiation. THz pulses with free-space field amplitudes of 500 kV/cm irradiated the sample at a 1-kHz repetition rate for periods of 2 seconds, separated by 5-minute intervals. Each emitted PL photon was detected by an avalanche photodiode and its arrival time relative to the photoexcitation pulse that preceded it was recorded with a time-to-digital converter (see Methods section). Every second, the recorded times were binned to yield an arrival time distribution. The distributions for a total temporal range of 32 minutes, including six 2-second THz irradiation periods and the 5-minute periods in between them, are shown in Fig.~4a. The data show a steep increase in the PL intensity when the THz excitation is turned on followed by a rapid and then slower decay in PL intensity when the THz excitation is turned off. An increase in PL photon counts beyond 10 ns is also observed during the same periods. As shown in Fig.~4b, the decays are non-exponential, indicating that multiple decay processes are observed. This is typical in thick-shell QDs due to charge separation and diffusion across the core/shell interface.\cite{Brovelli2011,Nan2012,Bodunov2018} The initial parts of the decays can be fit approximately to exponential forms with decay times of 2.4 ns and 3.2 ns with and without THz irradiation, respectively. We ascribe this difference to the action of the THz field in suppressing part of the fast non-radiative recombination responsible for the PL quench, consistent with the discharging of the QDs. This elongation is readily apparent, an increase of about 33\%, though the time scale is much closer to that of the quenched state than the long-lived state on glass. Although this change seems incongruous with the large change observed in the THz-induced PL intensity (which can increase many-fold) it is important to note that the PL intensity is primarily driven by the fraction of exciton emission, while the lifetime is dominated by the fastest relaxation process (\textit{e.g.}, trion decay). And the THz field does not neutralize quenched QDs entirely. Therefore, we do not expect any dramatic change upon THz irradiation in the early PL decay kinetics, during which the trionic and other fast decay channels dominate. At longer times, the dynamics do not return to the typical behavior observed in QDs on glass. This suggests that, although THz irradiation discharges QDs, the QDs are still in a different state from the neutral state of QDs on glass.

We also measured the PL emission spectra of QDs on gold with and without THz excitation. The results are shown in Fig.~4c. Upon a 2-second exposure to THz pulses at 50 s, the observed PL brightens significantly and then slowly decays thereafter. The averaged spectra before and after THz irradiation are shown in Fig.~4d. Before THz irradiation, the PL spectrum is peaked at around 621 nm and appears slightly asymmetric with a shoulder on the red side of the main peak. Following THz irradiation, we see a modest (about 3 nm) blue-shift in the spectral peak and a notable relative spectral weight transfer from the red shoulder to the main peak. For comparison, we measured the PL spectra of QDs on gold and glass substrates without any THz irradiation, shown in Fig.~4d and Fig.~S11. We observe a spectral shift of about 5 nm between the quenched and unquenched QD PL spectra on gold and glass, which is comparable to the 3-nm blue shift induced by THz pulses. We argue that this similarity arises from the same physical origin, \textit{i.e.}, charging. When excess charges are present, further irradiation creates a trion with low emission probability, quenching and redshifting the PL.\cite{Patton2003,Zhang2014,Bracker2005} The excess charges can also create an internal electric field and induce the quantum-confined Stark effect\cite{Empedocles1997}, which distorts the electron and hole wavefunctions and consequently redshifts the emission spectrum (detailed discussion is included in Supplementary Note 8). The blue-shifted spectra of QDs on gold after THz irradiation and QDs on glass suggest that both spectra emerge primarily from excitonic states without excess charges present. This further indicates that THz fields are able to effectively suppress charging-induced Auger recombination in QDs. 

\section{Conclusions}

Our results demonstrate a route to brightening quenched QDs, overcoming the limitations posed by the presence of excess charges. This, in turn, establishes an active control of non-radiative decay using short THz pulses. This effect may be broadly relevant to other luminescent materials for LED applications, \textit{e.g.}, organic compounds\cite{Burin2000,Han2012} and two-dimensional transition metal dichalcogenides\cite{Bhanu2014,Wang2018}, since quenching due to the presence of CTLs or other conductive surfaces is ubiquitous in these material systems and often limits device brightness and efficiency. Incorporation of THz waveguides in devices may simplify the implementation of this control scheme\cite{Ishikawa2009} for practical applications.

\clearpage
\newpage

\section{Methods}

\noindent \textit{THz metamaterial fabrication and enhancement simulation}\\
The fabrication of the metal microslit array is based on a standard photolithography and lift-off process. Image reversal photoresist AZ5214 was spin-coated on a fused silica substrate at 3000 rpm for 30 s, soft baked at 110 $^{\circ}$C for 50 s on a hotplate, UV exposed by a maskless aligner MLA 150 with a dose of 24 mJ/cm$^2$, and post-exposure baked at 120 $^{\circ}$C for 2 min followed by flood exposure and development in AZ422. A thin film of 10 nm Cr was deposited onto the substrate as an adhesion layer by thermal evaporation followed by a 90-nm-thick Au thin film. The sample was soaked in acetone for lift-off, rinsed by IPA, dried by nitrogen and gently cleaned by oxygen plasma to remove organic residue that may left on the sample. The non-uniformity of the gold edges can be found in the scanning electron microscope images reported in this literature (same fabrication method adopted)\cite{Pein2017}. For the oxide-coated metal microslit array, a 50-nm Al$_2$O$_3$ layer was deposited on top of the metal microslit array by atomic layer deposition. Simulations of microslit THz near-field enhancement have been presented\cite{Pein2017}. The enhancement has a spatial variation in the perpendicular direction to the gap due to the nature of the local enhancement of a capacitive gap as described previously\cite{Seo2009}. The sub-micron gold edge roughness may further add to the nonuniformity in the THz enhancement.\\

\noindent \textit{CdSe/CdS QDs synthesis}\\
To synthesize CdSe/CdS QDs, an established synthetic protocol from the literature was used.\cite{Carbone2007} To synthesize CdSe core QDs, 60 mg CdO, 280 mg octadecylphosphonic acid and 3 g trioctylphosphine oxide were combined in a 50 mL round bottom flask. The resulting reaction mixture was put under vacuum and heated to 150 $^{\circ}$C to remove volatile substances. After 1 hour, the mixture was heated further to 320 $^{\circ}$C under nitrogen flow to form a clear colorless solution, and 1.0 mL trioctylphosphine was added dropwise. The temperature was increased to 380 $^{\circ}$C and the heating mantle was removed. 0.5 mL Se/trioctylphosphine (60 mg Se in 0.5 mL trioctylphosphine) was injected rapidly and the reaction mixture was cooled down to room temperature with air. The crude reaction mixture was washed with acetone and redispersed in toluene. This synthetic protocol resulted in highly monodisperse core-only QDs  with first exciton absorption at 487 nm. To overcoat CdSe core QDs with CdS shells, continuous injection synthesis was used.\cite{Chen2013} In a 100 ml round bottom flask, 100 nmol CdSe core QDs in toluene, 3 mL octadecene (ODE), 3 mL oleylamine and 3 mL oleic acid were mixed. The resulting mixture was degassed for 20 minutes at room temperature and for 40 minutes at 100 $^{\circ}$C to remove volatile substances. Afterwards, the mixture was put under nitrogen flow, and the temperature was increased. When the temperature reached 200 $^{\circ}$C, 0.08 M cadmium oleate-ODE and octanethiol-ODE (1.2 equivalents) were added at 2.5 mL/hour to form 10 monolayers of CdS shell. After the completion of shell precursor injection, the reaction mixture was annealed at 310 $^{\circ}$C for 15 minutes. QDs were washed with acetone and redispersed in hexane three times. Finally, the solution is diluted with toluene and spin-coated on microslits to achieve a sparse distribution.\\

\noindent \textit{Theoretical modeling of unquenching of QD PL by THz fields}\\
In the model presented in Fig.~2g, we treat the state of the free hole with a three-state model with four variable rate constant parameters. The model is reversible and obeys detailed balance. The rate matrix $R$ is given by,
\begin{equation}
	R = 
	\begin{pmatrix}
		-k_{Q \rightarrow D} & -k_{D \rightarrow Q} & 0 \\
		k_{Q \rightarrow D} & -k_{D \rightarrow Q}-k_{D \rightarrow M} & k_{M \rightarrow D} \\
		0 & -k_{D \rightarrow M} & -k_{M \rightarrow D} 
	\end{pmatrix}
\end{equation}
This rate matrix can itself exist in one of three states: initial, response or relaxation. The initial and relaxation rate matrices are fit to reproduce stationary PL levels observed before the THz field is turned on ($t$ = -10 s to $t$ = 0 s). PL levels are taken to be proportional to the population of the $M$ state. Only the initial rate matrix of the first run is unique. In subsequent runs the initial rate matrix is the same as the relaxation matrix for the previous run (\textit{i.e.} $R_{2,initial}=R_{1,relax}$). The response rate matrix was kept universal across all four runs, such that the memory effects observed in the THz-on phase ($t$ = 0 s to $t$ = 10 s) can uniquely be attributed to varying initial populations between each run. The parameters for the response matrix were chosen to qualitatively reproduce the experimental data and are given in Table 1 of Supplementary Materials. Here we provide a qualitative justification for our choice of parameters, and a physical explanation for the behavior the model produces.

At the start of the experiment, the hole population almost entirely localized in the bulk quantum dot ($Q$ state) since the CdS shell valence band is well below the Fermi level of the metal. When the pulsed THz field is turned on, the rate matrix shifts instantaneously as a result of the energization of the free holes on the QDs. Relative to the initial rate matrix, the equilibrium between the $Q$ and $D$ states is shifted toward the $D$ state, and the equilibrium between the $D$ and $M$ states is shifted toward the $M$ state. However, the rate constants for transitions between the $Q$ and $D$ states are assumed to be an order of magnitude lower than those for the transitions between $D$ and $M$ states, making the response to the field highly dependent on the initial population of the $D$ state. In the first two runs, the initial population of the $D$ state is equal to or lower than that of the $Q$ state, leading to a slow rise in the population of the $M$ state, and a more gradual PL response (see Fig.~S10).  However in the final two runs, the $D$ state is initially the most populated, leading to a much faster rise in $M$ when the THz pulses are turned on.

While this simple kinetic model qualitatively captures the discrepancies in the responses to the THz field observed between runs, it is worth noting that it does not accurately model the temporal profiles of the PL curves, particularly for runs 3 and 4. One may try to fit the parameters of $R_{response}$ to better agree with Runs 3 and 4, but this results in a less accurate representation of Runs 1 and 2 (Fig.~S9). Introducing multiple distinct $D$ states, perhaps to account for the heterogeneity of surface defects, may give our model enough flexibility to accurately capture the behavior of all four runs. However, for the purposes of this study, we believe it was best to utilize a simpler model with fewer free parameters. We also acknowledge that our model does not given us the correct power-law decay behavior, and requires different relaxation rate matrices for each run. In future studies, we hope to enhance the model to help understand such behavior by considering the role of biexciton states, as these states have been shown to play a significant role in the power-law behavior of QD blinking.\cite{efros_random_1997,peterson_modified_2009,ye_blinking_2011,efros_origin_2016}\\

\noindent \textit{High-field THz pulse generation}\\
High-field THz pulses were generated in a Mg:LiNbO$_3$ (LN) crystal by tilting the optical pulse front to achieve phase matching.\cite{Hebling2008}. By using a two-parabolic-mirror terahertz imaging system, the image of the THz spot on the sample was reduced to its diffraction limit of 600 $\mu$m in diameter. The incident THz field temporal profile was measured in the time domain using electro-optic sampling\cite{Wu1995} with a 100 $\mu$m thick 110-oriented gallium phosphide crystal. When pumping the LN crystal with an amplified Ti:sapphire laser system output (repetition rate 1 kHz, central wavelength 800 nm, pulse duration 35 fs, output power 12 W), the maximum electric field of the THz pulses reached 700 kV/cm at the focus, with a spectrum centered at 0.65 THz. The field was attenuated to be 500 kV/cm to maintain reversibility of the measurement. THz repetition rates were down counted from 1 kHz to 500 Hz, 250 Hz, 125 Hz and 62.5 Hz by using four successive optical choppers.\\

\noindent \textit{PL measurements}\\
For PL imaging, a continuous-wave laser beam (center wavelength 405 nm, Toptica iBeam smart) was used to illuminate the sample with an excitation intensity up to 150 W/cm$^2$. The PL image was collected by a 100$\times$ objective with a numerical aperture of 0.7 and imaged on an sCMOS camera (Andor Zyla 5.5). The zero of time was defined by the opening of a shutter, and successive PL images were recorded every 0.3 sec. For PL spectrum characterization (also on a time scale of seconds), a confocal geometry was used to select an area on one field enhancement gap edge of the THz metamaterial. Successive measurements of the spectrum were recorded in a spectrometer (Andor Shamrock spectrometer and Andor Newton 920) at 1 sec intervals. For PL lifetime measurements with 420 ps time resolution (Instrumental response function is provided in Fig.~S8), a pulsed laser source (center wavelength 405 nm, repetition rate 5 MHz, 100 ps pulse duration, Picoquant) was used to excite the sample with an excitation intensity around of 2 $\mu$J/cm$^2$. The PL signal was collected by the same objective and confocal microscope geometry described above and detected by an avalanche photodiode (ID Quantique, id100-vis). The lifetime was obtained by tagging the arrival time of each PL photon relative to the trigger photon using a time-to-digital converter (ID Quantique, id800-TDC). The duration and repetition rate of THz irradiation were controlled by a mechanical shutter. For the error bar estimation, videos/image series were obtained. By calculating the time-averaged PL intensity and deviation with and without THz excitation, the mean and standard deviation values of THz-induced PL intensity changes were determined. All the above measurements were performed in an ambient environment.


\begin{acknowledgement}
We acknowledge technical support and helpful discussions from Lili Wang and Weiwei Sun. J.S., F.Y.G., Z.Z., H.U., U.B., A.F., A.P.W., M.G.B., and K.A.N. acknowledge supported in part by the U.S. Army Research Lab (ARL) and the U.S. Army Research Office through the institute for Soldier Nanotechnologies, under Cooperative Agreement number W911-NF-18-2-0048. J.S., F.Y.G., Z.Z., and K.A.N. acknowledge additional support from the Samsung Global Outreach Program. N.G and E.B. acknowledge support from the US Department of Energy, BES DMSE, Award number DE-FG02-08ER46521. E.B acknowledges additional support from the Swiss National Science Foundation under fellowships P2ELP2-172290 and P400P2-183842. J.H. and  V.B. acknowledge support from the Center for Energy Efficient Electronics Science (NSF Award 0939514). J.D. acknowledges financial support from the NSERC Postgraduate Scholarships program.
\end{acknowledgement}

\section{Author Contributions} $^\bigtriangleup$ J.S., F.Y.G., and Z.Z. contributed equally to this work. K.A.N., J.S., F.Y.G., Z.Z., and H.U. conceived the idea. J.S., F.Y.G., and Z.Z. performed all the measurements and analyzed the data with support from H.U., U.B., C.-W.B., and K.S.C.; U.B. synthesized CdSe/CdS quantum dots under the supervision of M.G.B.; A.F. performed kinetic model simulation under the supervision of A.P.W; J.H. fabricated THz metamaterial structures under the supervision of V.B.; J.S., F.Y.G., Z.Z., E.B., and K.A.N. wrote the manuscript with inputs from all the authors. K.A.N. supervised the project.

\section{Notes}
The authors declare no competing financial interest. Correspondence and requests for materials should be addressed to K.A.N. (email: kanelson@mit.edu).
	
\begin{suppinfo}
	Supplementary information is available in the online version of the paper.
\end{suppinfo}

\begin{figure}[t]
	\begin{center}
		\includegraphics[width=\columnwidth]{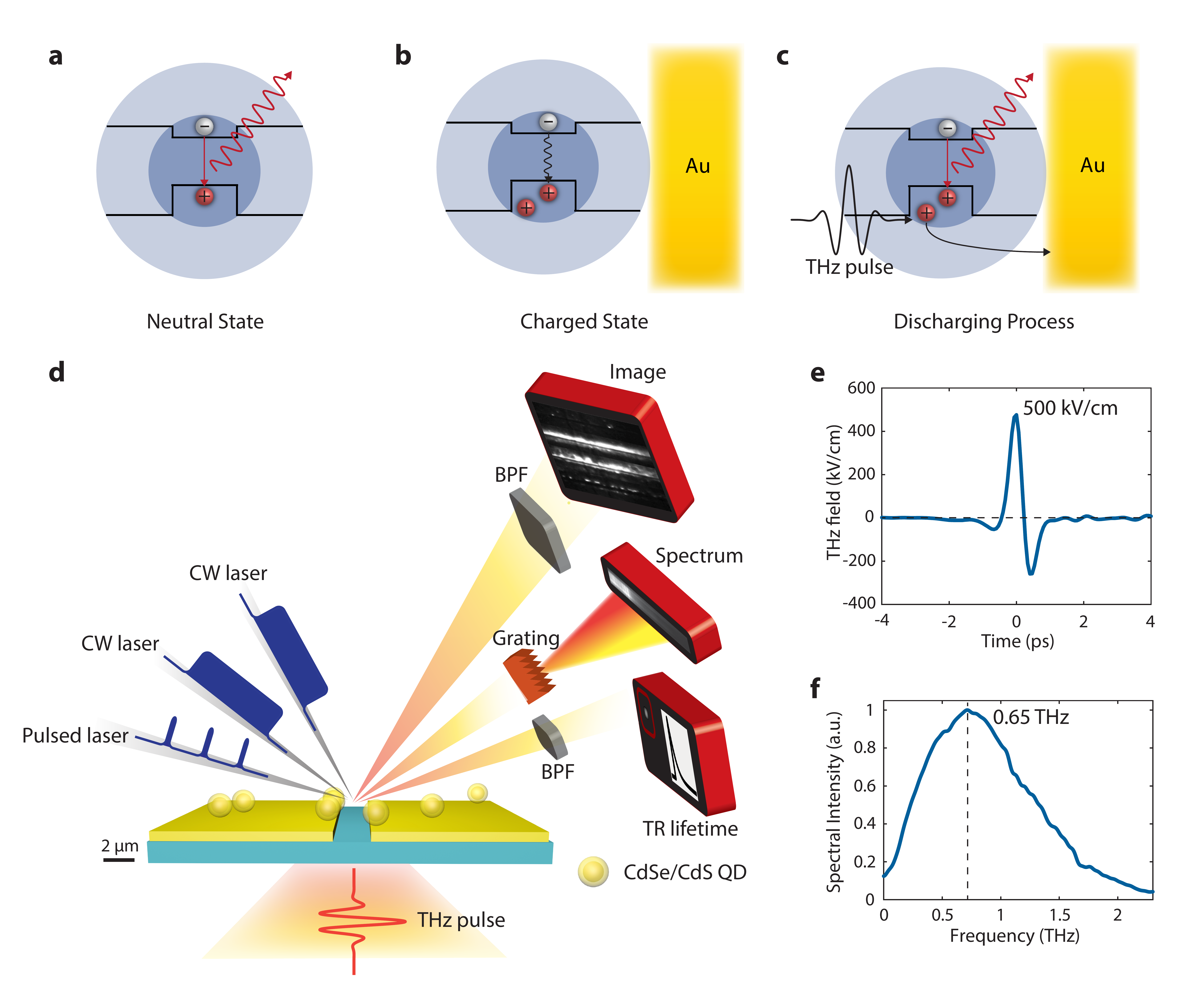}
		\caption{\textbf{THz brightening of suppressed PL of CdSe/CdS QDs on gold.} (a) Neutral QDs exhibit high PL quantum yield through excitonic emissions. (b) When placing QDs in contact with gold, optically generated excitons break into separate charges, with the electron transferring to gold. This leaves an excess hole in the QD, which leads to weakly radiative trion recombination and non-radiative relaxation and quenches the PL. (c) The THz field accelerates the free hole in the QD, and it reaches energies that exceed the interfacial barrier between QDs and gold, resulting in the removal of the hole and the unquenching of the PL. (d) Schematic illustration of sparse QD distributions on the THz metamaterial structure with THz pulse excitation and various PL probes, including PL imaging, PL spectrum measurement, and (with pulsed rather than CW photoexcitation) time-resolved (TR) PL lifetime measurements. (e) The free-space THz pulse has a peak electric field strength of 500 kV/cm. (f) The free-space THz spectrum is centered around a frequency of 0.65 THz.}
		\label{fig:Fig1}
	\end{center}
\end{figure}

\begin{figure}[t]
	\begin{center}
		\includegraphics[width=\columnwidth]{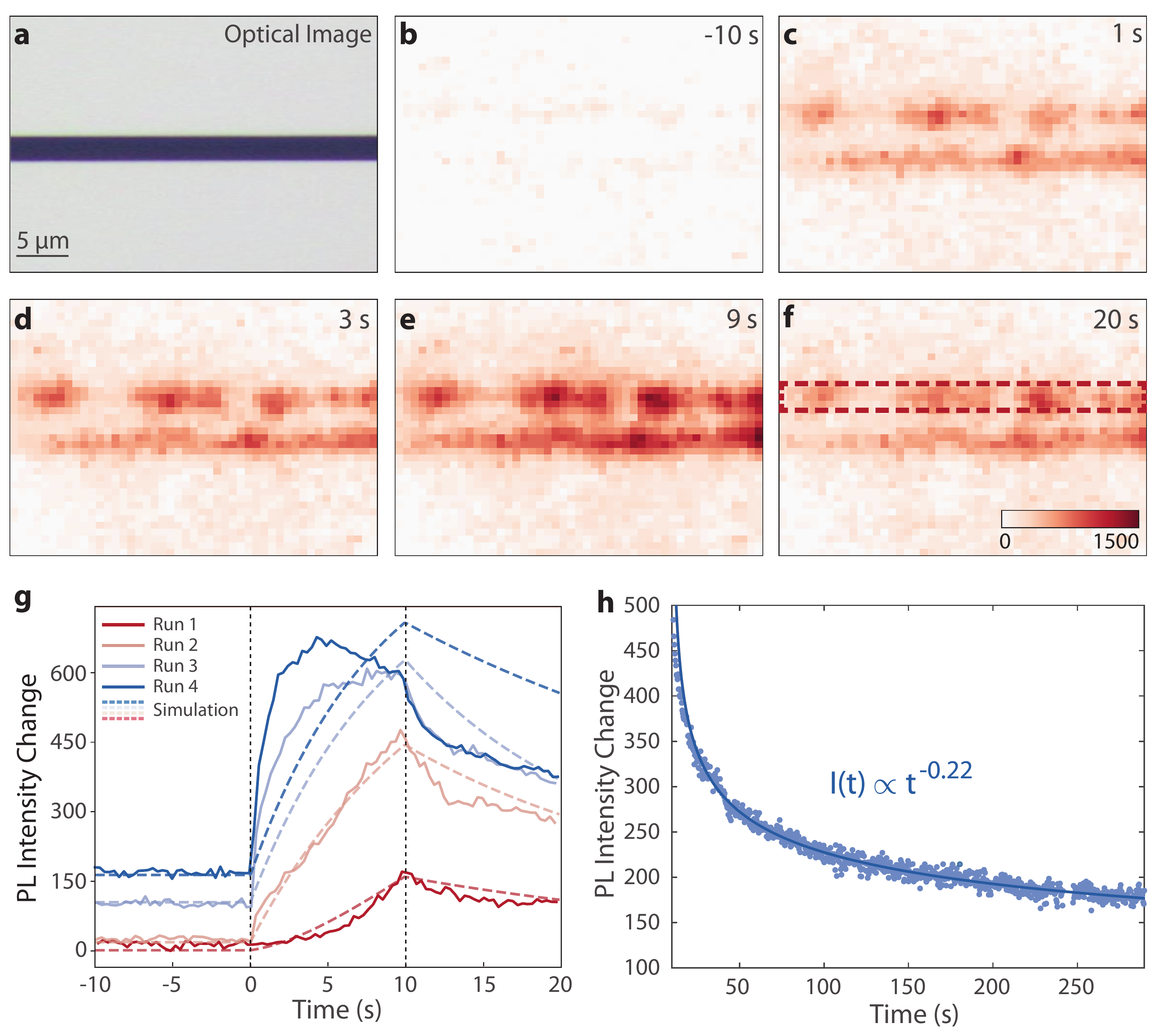}
		\caption{\textbf{THz discharging dynamics.} (a) An optical image of a 2-$\mu$m wide insulating gap (the dark horizontal line) between 100-$\mu$m wide gold strips in a THz field enhancement structure. THz field enhancement occurs in and around the gap, and the highest field enhancement occurs around the gold immediately adjacent to the gap. (b-f) Images show the PL intensity from an identical THz metamaterial region with sparsely distributed QDs on top, under CW optical irradiation. The scale bar is the same as (a). Before THz irradiation ($t$ < 0), PL from the sparse QDs is almost completely quenched. When THz pulses at a 1-kHz repetition rate are turned on by the opening of a shutter at time zero, PL emission appears near the edges of the insulating gap and evolves over the next several seconds. Even tens of seconds after the THz pulses are blocked at $t$ = 10 s, PL intensity can still be observed from locations where there was none before THz excitation. }
		\label{fig:Fig2}
	\end{center}
\end{figure}
\addtocounter{figure}{-1}
\begin{figure} [t]
	\caption{(g) The temporal evolution of the PL intensity (integrated spatially over the upper edge of the metamaterial insulating gap, outlined in red in Fig.~2f) for a set of four THz excitation experiments conducted on the same slit from the same sample. Measurements were recorded for 10 s before, during, and after THz irradiation with the dashed lines demarcating the window for THz exposure. An interval of 300 s separated successive runs. PL intensity increases slowly during THz excitation in the first run but increases more rapidly, reaching higher peaks, in the subsequent runs. The data from different runs are not offset vertically in the figure; there is significant PL emission prior to runs 3 and 4, even after the 5-minute interval between runs. The lighter dashed lines show the corresponding results from a kinetic model that simulates the dynamics of charge transfer between the QDs and the gold due to the THz field. The image sequence shown (b-f) is from Run 3 of this sample. (h) The quenching recovery process after Run 4 plotted for times after the THz beam is turned off. The recovery of quenching is not complete even after 5 minutes. The long-time kinetics are well described by a power-law decay with an exponent of 0.22.}
\end{figure}

\begin{figure}[t]
	\begin{center}
		\includegraphics[width=\columnwidth]{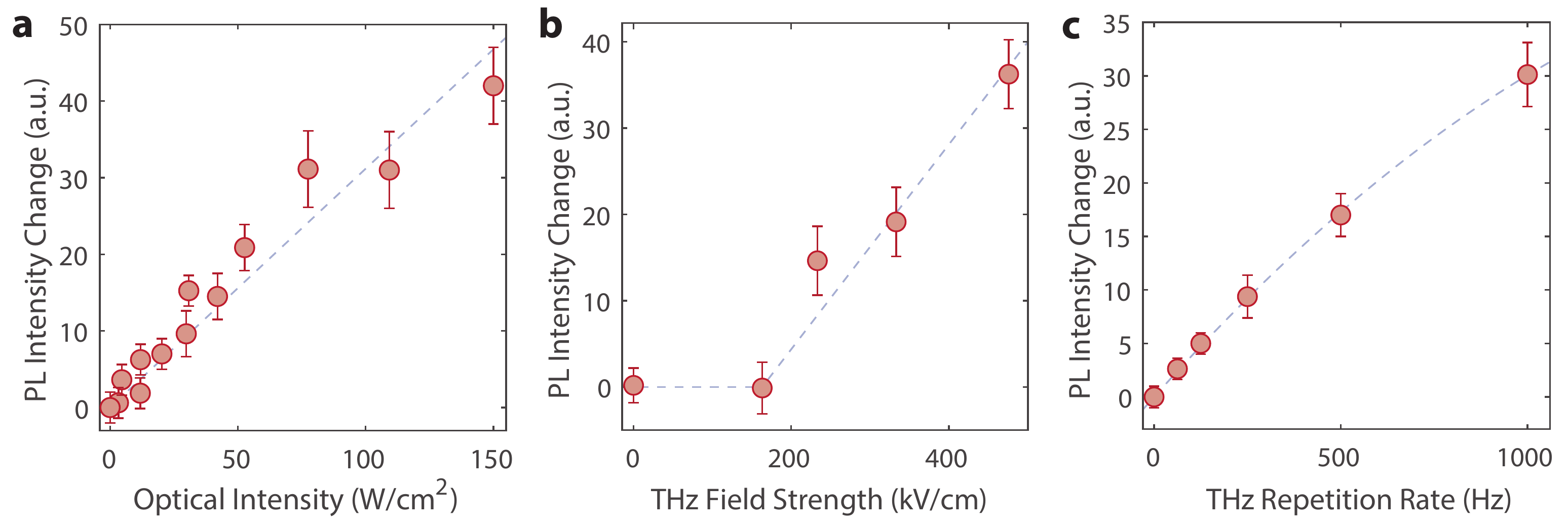}
		\caption{\textbf{THz-induced PL intensity dependences on optical excitation intensity, THz field strength, and THz repetition rate.} All measurements were conducted with CW 405-nm excitation. The PL intensities reported were integrated over 2-second periods of THz irradiation at 1-kHz repetition rate. Successive measurements were separated by 20 minutes to allow adequate time for sample recovery. PL signal was measured from a region near the gold gap edge, similar to that outlined in Fig.~2f. The same region of the same sample was used in all measurements presented in this figure. The dashed lines serve as guides for the eye. (a) CW optical intensity-dependence of the THz brightening of QD PL intensity. (See supplementary information for images). The THz-induced PL signal change scales roughly linearly with optical power. No luminescence is observed without optical excitation. (b) THz field strength-dependence. No obvious effect is observed when the incident field strength is lower than $\sim$ 170 kV/cm. (c) THz repetition rate-dependence. The THz-induced PL intensity change is approximately proportional to the THz repetition rate, \textit{i.e.}, to the total number of THz pulses that irradiated the sample during each 2-second irradiation period.}
		\label{fig:Fig3}
	\end{center}
\end{figure}

\begin{figure}[t]
	\begin{center}
		\includegraphics[width=\columnwidth]{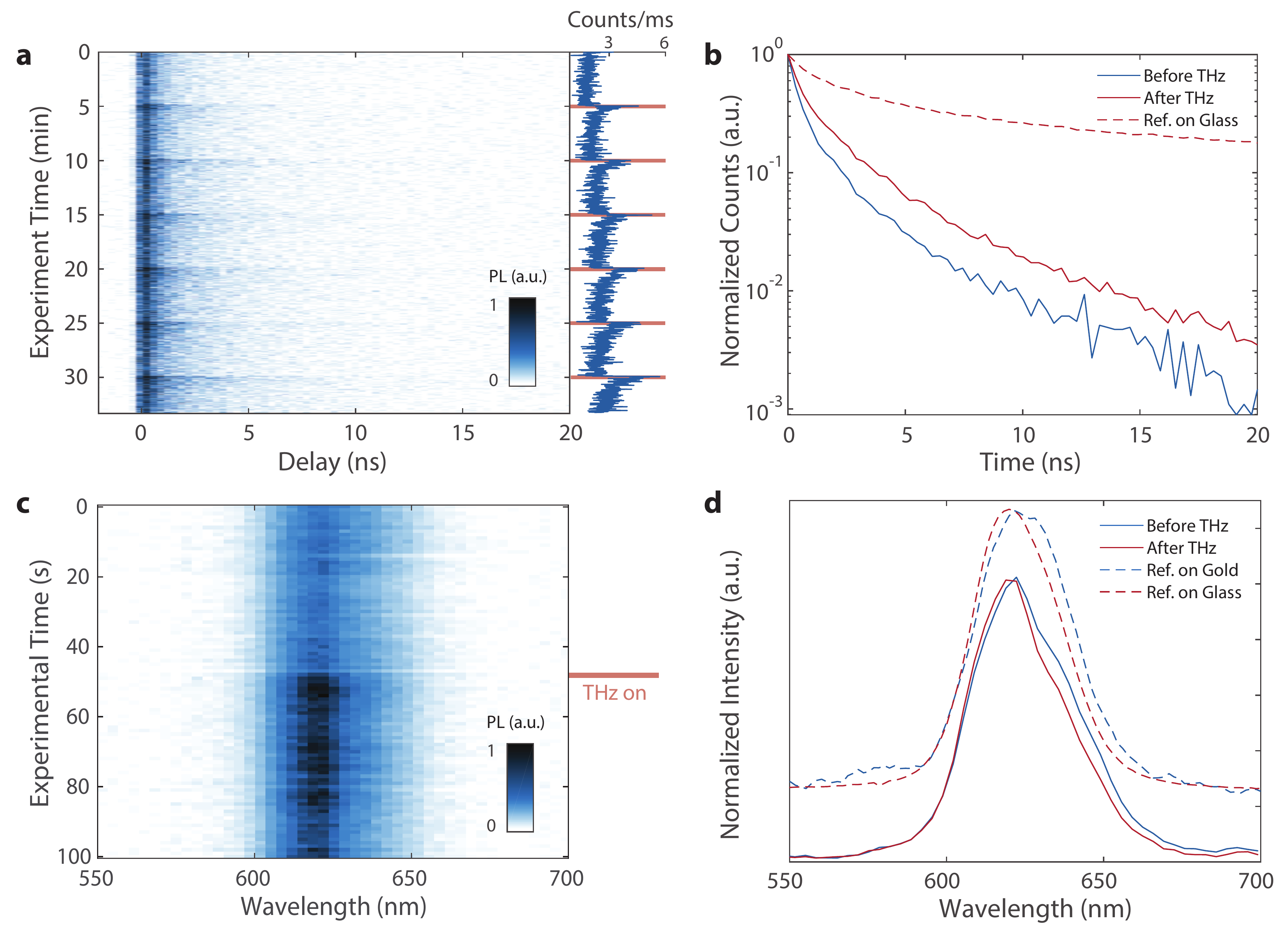}
		\caption{\textbf{PL kinetics and PL spectrum measurements.} (a) Time-tagged PL photon counts over each second are shown as a function of experimental time with the total counts displayed to the right. The 2 seconds of THz irradiation every 5 minutes result in a spike in the overall PL that decays rapidly after the THz irradiation stops and then continues to decay slowly over the next few minutes. (b) Averaged time-resolved PL decay curves corresponding to THz off and THz on times show an increase in the lifetime from 2.4 ns to 3.2 ns during THz irradiation. The lifetime distributions for the same QDs on glass are also shown for reference. (c) The PL emission spectra before, during and after a 2 s window of THz exposure. An increase in PL intensity can be seen following THz exposure. (d) Normalized averaged spectra over experimental time from before and after THz exposure show a small (3-nm) blue shift similar to the slightly larger blueshift in the emission spectra from QDs on glass relative to QDs on gold (shifted vertically for clarity).}
		\label{fig:Fig4}
	\end{center}
\end{figure}

\clearpage
\newpage


\providecommand{\noopsort}[1]{}\providecommand{\singleletter}[1]{#1}%
\providecommand{\latin}[1]{#1}
\makeatletter
\providecommand{\doi}
{\begingroup\let\do\@makeother\dospecials
	\catcode`\{=1 \catcode`\}=2 \doi@aux}
\providecommand{\doi@aux}[1]{\endgroup\texttt{#1}}
\makeatother
\providecommand*\mcitethebibliography{\thebibliography}
\csname @ifundefined\endcsname{endmcitethebibliography}
{\let\endmcitethebibliography\endthebibliography}{}

\newpage
\clearpage

\end{document}